\documentclass[12pt,showkeys,pra,superscriptaddress,floatfix,nofootinbib,longbibliography]{revtex4-2}

\usepackage{graphicx}
\usepackage{color}
\usepackage{amsfonts}
\usepackage{eurosym}
\usepackage{xspace}
\usepackage{ulem}
\renewcommand{\d}{{\rm d}}
\renewcommand {\i} {{\rm i}}

\newcommand {\E}  {\varepsilon}
\newcommand {\om} {\omega}

\newcommand {\calA} {{\cal A}}

\begin{document}

%%%%%%%
\title{Construction of crystal-based light sources with LWFA
electron beams}

\author{Illia Zymak}
\email[]{Illia.Zymak@eli-laser.eu}
\affiliation{ELI Beamlines, The Extreme Light Infrastructure ERIC, Dolní B\v{r}e\v{z}any, Czech Republic}

\author{Michal Nevrkla}
%\email[]{\textcolor{red}{NameOne@eli-laser.eu}}
\affiliation{ELI Beamlines, The Extreme Light Infrastructure ERIC, Dolní B\v{r}e\v{z}any, Czech Republic}

% \author{Gabriele Grittani}
% %\email[]{\textcolor{red}{NameTwo@eli-laser.eu}}
% \affiliation{ELI Beamlines, The Extreme Light Infrastructure ERIC, Dolní B\v{r}e\v{z}any, Czech Republic}

\author{Ji\v{r}i \v{S}i\v{s}ma}
%\email[]{\textcolor{red}{NameTwo@eli-laser.eu}}
\affiliation{ELI Beamlines, The Extreme Light Infrastructure ERIC, Dolní B\v{r}e\v{z}any, Czech Republic}

\author{Filip Vitha}
%\email[]{\textcolor{red}{NameTwo@eli-laser.eu}}
\affiliation{ELI Beamlines, The Extreme Light Infrastructure ERIC, Dolní B\v{r}e\v{z}any, Czech Republic}

%%%%%%%
\author{Nigel J. Mason}
%\orcid{0000-0002-4468-8324}
\email{n.j.mason@kent.ac.uk}
\affiliation{Physics and Astronomy,
School of Engineering, Mathematics and Physics, University of Kent,
Park Wood Rd, Canterbury CT2 7NH, UK}

%%%%%%%
\author{Maykel M\'arquez-Mijares}
%\orcid{0000-0003-2503-2802}
%\email[]{mmarquez@instec.cu}
\affiliation{Instituto Superior de Tecnolog\'ias y Ciencias Aplicadas,
University of Havana (InSTEC-UH), Ave. Salvador Allende 1110, Plaza de
la Revoluci\'on, Havana - 10400, Cuba}

\author{Germ\'an Rojas-Lorenzo}
%\orcid{0000-0002-7605-9426}
%\email[]{german@instec.cu}
\affiliation{Instituto Superior de Tecnolog\'ias y Ciencias Aplicadas,
University of Havana (InSTEC-UH), Ave. Salvador Allende 1110, Plaza de
la Revoluci\'on, Havana - 10400, Cuba}

\author{Jes\'us Rubayo-Soneira}
%\orcid{0000-0001-6093-9181}
%\email[]{jrubayo@gmail.com}
\affiliation{Instituto Superior de Tecnolog\'ias y Ciencias Aplicadas,
University of Havana (InSTEC-UH), Ave. Salvador Allende 1110, Plaza de
la Revoluci\'on, Havana - 10400, Cuba}
% \author{Gennady B. Sushko}
% % \email[]{}
% \affiliation{MBN Research Center, Altenh\"{o}ferallee 3, 60438 Frankfurt am Main, Germany}

\author{Andrei V. Korol}
\email[]{korol@mbnresearch.com}
\affiliation{MBN Research Center, Altenh\"{o}ferallee 3, 60438 Frankfurt am Main, Germany}

\author{Andrey V. Solov'yov}
\email[]{solovyov@mbnresearch.com}
\affiliation{MBN Research Center, Altenh\"{o}ferallee 3, 60438 Frankfurt am Main, Germany}

%%%%%%%%%%%%%%%%
\begin{abstract}
We discuss a possibility to construct a crystal-based light source (CLS)
of gamma-rays using electron beams produced by  Laser Wake
Field Acceleration (LWFA).
This state-of-the-art acceleration technique is capable of producing
few-GeV, ultra-short (fs) electron beams.
In the case studies presented, we demonstrate the feasibility of
novel gamma-ray CLSs based on exposing single crystals to
electron beams available at the ELI-ELBA.
The results of rigorous numerical simulations are presented
to  characterise the radiation emitted emitted by
1 and 4 GeV electrons incident on a 200-micron-thick
single silicon crystal along the (110) plane.
The simulations were performed using the current divergence
as well as a smaller value that is believed to be a realistic
to achieve in the near future.
We demonstrate that it is feasible to achieve a photon yield
of $10^6-10^7$ photons/s in the photon energy range from 5
to 50 MeV.
Looking further ahead, we also performed a similar analysis
for a 4-GeV positron beam incident on much thicker targets, up to
2 mm oriented silicon crystal.
In this case, the photon yield can be increased by an order of
magnitude.
\end{abstract}

% \pacs{61.85.+p, 41.60.-m, 41.75.Ht, 02.70.Uu, 07.85.Fv}

%61.85.+p  Channeling phenomena
%41.60.-m  Radiation by moving charges} \and
%41.75.Ht  Relativistic electron and positron beams} \and
%02.70.Uu  Applications of Monte Carlo methods}}
%07.85.Fv X- and γ-ray sources, mirrors, gratings, and detect
\maketitle

 %%%%%%%%%%%%%%%%%%%%%%%%%%%%%%%%%%%%%%%%%%%%%%%%%%%%%%%%%%%%%%%%%
\section{Introduction and motivation  \label{Introduction}}

One of the approaches developed to generate intense quasi-monoenergetic
gamma rays with variable photon energies relies on the laser-Compton
scattering phenomenon.
In this process, a low-energy laser photon is scattered backwards by an
ultra-relativistic electron, acquiring an energy increase proportional
to the electron's squared Lorentz factor, $\gamma \gg 1$
\cite{Federici-EtAl:NuovoCim_vol59B_247_1980}.
This method has been used to produce gamma rays over a broad energy
range.
Reviews of the commissioned and under-construction facilities of the inverse Compton
scattering gamma-ray light sources (LS)  are available in Refs.
\cite{AlbertThomas_PlasmaPhysContrFusion_v58_103001_2016,
Krafft-Priebe:RevAccScieTechnol_vol3_p147_2010,
Muramatsu-EtAl:NIMA_v737_p184_2014,
Sei-EtAl:ApplSci_vol10_p1418_2020,NextGenerationGammaRayLS2022,
Wu-EtAl:PRL_vol96_224801_2006}.

Intense gamma rays can be generated using crystal-based LS (CLS),
which is set up by exposing oriented crystals of
different shapes and orientations to ultrarelativistic electron
and positron beams
\cite{KorolSolovyov:EPJD_CLS_2020,CLS-book_2022,%
KorolSushkoSolovyov:PRAB_v27_p100703_2024}.
These novel gamma-ray LS can operate in the
photon energy range from hundreds of keV to the GeV domain
\cite{CLS-book_2022}, which is
an area of current interest in many fields including fundamental
science, industry, biology and medicine
\cite{AlbertThomas_PlasmaPhysContrFusion_v58_103001_2016,%
Rehman_EtAl_ANE_v105_p150_2017,Kraemer_EtAl-ScieRep_v8_p139_2018,
NextGenerationGammaRayLS2022,BudkerEtAl:AnnPhys_v534_2100284_2022}.
The practical realisation of CLSs is the primary objective of
the current  Horizon  Europe EIC-Pathfinder-2021
project TECHNO-CLS \cite{TECHNO-CLS}.

As was recently demonstrated by means of rigorous numerical
simulations, the intensity of radiation from CLS
can exceed that achievable at modern laser-Compton facilities
\cite{KorolSushkoSolovyov:PRAB_v27_p100703_2024}.
In particular, it was shown that for moderate average beam
currents up to 10 $\mu$A, the average photon flux in
the energy range of $10^0-10^2$ MeV emitted within a cone of
$10^1-10^3$ $\mu$rad and 1 per cent of the bandwidth,
can be as high as $10^{10} - 10^{12}$ photon/s.
These values are higher than the fluxes available at modern
laser-Compton gamma-ray LSs, which operate with beam currents
orders of magnitude higher (see, for example, Table I in
\cite{KorolSushkoSolovyov:PRAB_v27_p100703_2024}).

The operational principle of CLSs relies on the phenomenon known as
channelling motion of electrons and positrons \cite{CLS-book_2022}.
This type of motion is experienced by ultra-relativistic charged
particles that are incident on oriented crystals at small angles to
the crystal plane (or axis), and that move further along the
crystallographic direction due to the collective action of the
electrostatic fields of the lattice atoms \cite{Lindhard}.
The crystalline fields can be as strong as $10^{10}$
V/cm, equivalent to a magnetic field of 3000 Tesla.
This value is two orders of magnitude greater than the 45.5 Tesla
record achieved in 2026 using superconducting magnets
\cite{Gao-EtAl:ScieAdv_vol12_eadz5826_2026}.
This enables the guided motion of projectiles in crystals of different
geometries and enhances the emitted radiation.
A channeling particle emits intense channeling radiation (ChR)
\cite{ChRad:Kumakhov1976,TerhunePantell:Kumakhov1976}, which is the
basis for CLS based on single crystals
\cite{KorolSushkoSolovyov:PRAB_v27_p100703_2024}.
In addition, the channeling motion in bent crystals gives
rise to intense synchrotron radiation the parameters of which
can be tuned by varying the bending radius within the range
$10^0-10^2$ cm \cite{Mazzolari-EtAl:PRL_112_135503_2014,
Shen-EtAl:NIMB_v424_p26_2018}.
In a periodically bent crystal, the crystalline undulator
radiation is emitted in a distinctly different range of
photon energies from ChR \cite{KSG_review2004,ChannelingBook2014}.
Its intensity, photon range, and peak width can be tuned by
changing the beam energy, the bending amplitude and period,
the type and thickness of the crystal, and the detector aperture
\cite{CLS-book_2022}.
The first experimental detection of the crystalline undulator
radiation has recently been reported
\cite{Backe-EtAl:PhysRevRes_v8_023200_2026}.

The operational efficiency of CLS relies on the high quality of the
crystalline target (including a very low level of defects) and the
incoming particle beam (with low emittance and energy spread).
Single crystals of diamond and silicon provide high structural quality
and have long been used in channeling and radiation experiments
involving ultra-relativistic electron and positron beams at various
accelerator facilities, including CERN, SLAC and the
Mainz Microtron (MAMI)
\cite{Uggerhoj:RPM_v77_p1131_2005,BiryukovChesnokovKotovBook,%
Wienands_EtAl-NIMB_v402_p11_2017,BackeEtAl:JINST_v13_C04022_2018,%
Yakimenko-EtAl:PR-AB_v22_101301_2019,%
Mazzolari-EtAl:PRL_v135_205002_2025,%
Agapov-EtAl:EPJPlus_v141_271_2026}.
The beams available at these facilities are of the required quality.
However, the main drawback of using these facilities for full-scale
research in the field of CLS and for the application of the produced
gamma radiation is the difficulty in obtaining sufficient beam time due
to high demand from a large number of other groups and collaborations.
Furthermore, several of these facilities are due to close for
refurbishment in next years and/or due to key staff retirements.

To overcome this drawback, it would be worthwhile looking for other
facilities that possess the necessary infrastructure to carry out
experiments with novel intense gamma-ray CLS.
In the future, these facilities could incorporate operational
CLSs into their infrastructure
and make them available for various applications.
In the course of the TECHNO-CLS project \cite{TECHNO-CLS}, a search
has been maintained on a permanent basis for a new beamline based
at an identified partner facility.

A group of such facilities has been identified that
houses laboratories employing electron
acceleration techniques based on Laser Wake Field Acceleration (LWFA)
with extremely high accelerating gradients ($> 100$ MeV/mm)
\cite{ELI-Ref01,ELI-Ref02,ELI-Ref03,ELI-Ref04}
and that have attained a state-of-the-art level.
These cutting-edge laboratories are therefore capable of delivering
ultra-short (fs range) high-energy electron
beams (GeV range) \cite{ELI-Ref14,ELI-Ref14a,ELI-Ref14b}.

The main goal of this paper is to demonstrate the feasibility of
novel gamma-ray CLSs based on exposing single crystals to
electron beams available at the ELI-ELBA facility
\cite{ELI-Ref14}.\footnote{When this manuscript has been finalised
for the submission, we become aware of Ref. \cite{Sytov-LWFA}, where
compact light sources that combine LWFA with oriented crystalline
targets were discussed.}
To achieve this, we performed rigorous numerical simulations of the
channeling efficiency and characterisation of the emitted radiation
for 1 and 4 GeV electrons incident on a 200-micron-thick
single silicon crystal along the (110) plane.
One of the key parameters used in the simulations was the angular
divergence of the beam.
The simulations were performed using the current divergence
as well as a smaller value that is realistic to achieve in the
near future.
We demonstrate that it is feasible to achieve a photon yield
of $10^6-10^7$ photons/s in the photon energy range from 5
to 50 MeV.
Looking further ahead, we also performed a similar analysis
for a 4-GeV positron beam incident on much thicker targets, up to
2 mm oriented silicon crystal.
In this case, the photon yield can be increased by an order of
magnitude.

This paper is organised as follows.
% \textcolor{red}{
% \sout{Section \ref{ELBA} provides an overview of the
% current status and outlook for the future development of
% using the ELBA beamline's multi-GeV electron
% (and positron?) beams to produce intense gamma radiation
% using CLSs.}
% }
An overview of the current capabilities and foreseen advancements
of the ELBA beamline is presented in Section II.
This section focuses on utilizing multi-GeV electron and possibly
positron beams to produce intense gamma radiation via CLSs.
Section \ref{Methodology} outlines the theoretical and computational
frameworks used in the numerical simulations.
In Section \ref{CaseStudies} we discuss the results of computational
characterisation of CLS based on single silicon crystals exposed to
the beams currently available at ELBA and to those which may
appear in the future.
A summary of the results obtained is given in the conclusions.

%%%%%%%%%%%%%%%%%%%%%%%%%%%%%%%%%%%%%%%%%%%%%%%%%%%%
\section{Present and future of ELBA beamline capabilities}
\label{ELBA}

% \textcolor{blue}{Highlighted in blue is input from ELI, received
% on 18.09.2026.}
% \textcolor{red}{This text should be modified further to include
% technical parameters of the existing beam (size,divergence) as well as
% the projections for the future.}

% \textcolor{blue}{
% The LWFA technique
% \cite{ELI-Ref01,ELI-Ref02,ELI-Ref03,ELI-Ref04}
%  has reached a stage of maturity where it can be put forward as a
%  reliable solution for steady hyperrelativistic beam delivery
%  technology encompassing a remarkable array of applications at
%  User facilities \cite{ELI-Ref05}.
% }

The LWFA technique
\cite{ELI-Ref01,ELI-Ref02,ELI-Ref03,ELI-Ref04}
has reached a stage of maturity where it can be put forward as a
reliable solution for steady ultrarelativistic beam delivery
technology encompassing a remarkable array of applications at
user facilities \cite{ELI-Ref05}.
Unlike conventional large-scale accelerator systems,
based on high-power RF fields, LWFA exploits the electric fields
of extreme amplitude generated by ultra-intense and ultra-short
laser pulses focused on gas targets, creating highly
nonlinear plasma for generating relativistic electron beams.

This method produces electron beams at much smaller scales and is
thus a compelling alternative acceleration technology.
The LWFA-based ALFA \cite{ELI-Ref06,ELI-Ref07,ELI-Ref08} and
ELBA \cite{ELI-Ref09,ELI-Ref10}   beamlines, alongside the
ELI Beamline Gammatron \cite{ELI-Ref11} are, at present, routinely
operated at The Extreme Light Infrastructure ERIC -- ELI
Beamlines Facility with proven several successful campaigns
involving several tens of users
for irradiation and quantum electrodynamics experiments,
and for driving secondary photon sources.

Applying this technique yields a few significant undisputed key
benefits, namely its high acceleration gradient (usually order of
magnitude for 1 GeV/cm) and therefore short acceleration length, high
bunch charge and natively ultra-short pulse duration (in the range
of tens of fs).
Nonetheless, parameters such as beam divergence and its pointing
stability as well as energy monochromaticity underperform those
found for conventional RF linear accelerators \cite{ELI-Ref12},
or storage devices \cite{ELI-Ref13}.
This results in increased effective beam emittance values and adds
complexity to the processes of photon generation using magnetic
undulators or inverse Compton scattering mechanisms.

Over the past years, the ELI Beamlines Electron Acceleration group
has steadily and successfully implemented several techniques, previously
exclusively available at laboratory level, to optimize both beam quality
and charge, namely through guided beam acceleration \cite{ELI-Ref14}
and nanoparticle-assisted electron injection \cite{ELI-Ref15} methods.
The first technique relies on the use of pre-formed plasma channels and
complementary laser beams that steer confined high-intensity laser pulses
over extended distances of the order of tens of cm to effectively
circumvent the focal spot Raleigh length limitation.
By overcoming this limitation, this approach achieves multi-GeV energy
ranges, whilest improving beam pointing stability and reducing divergence
to the sub-milliradian range.
The second technique, known as nanoparticle-assisted electron injection,
is a method where nanoparticles of a condensed medium are mixed with the
gas target.
This approach allows enhancement of the electron density within a given
injection volume, thereby increasing available electron bunch charge
while preserving the same laser and gas target configuration.
The simultaneous integration of these technologies at ELBA enables the
delivery of up to 5 GeV electron beams with angular charge density of
nC/mrad at a repetition rate of a few Hz, or in shot-on-demand regime.

% %%%%%% ELI, Figure
\begin{figure} [h]
\centering
\includegraphics[width=0.75\textwidth]{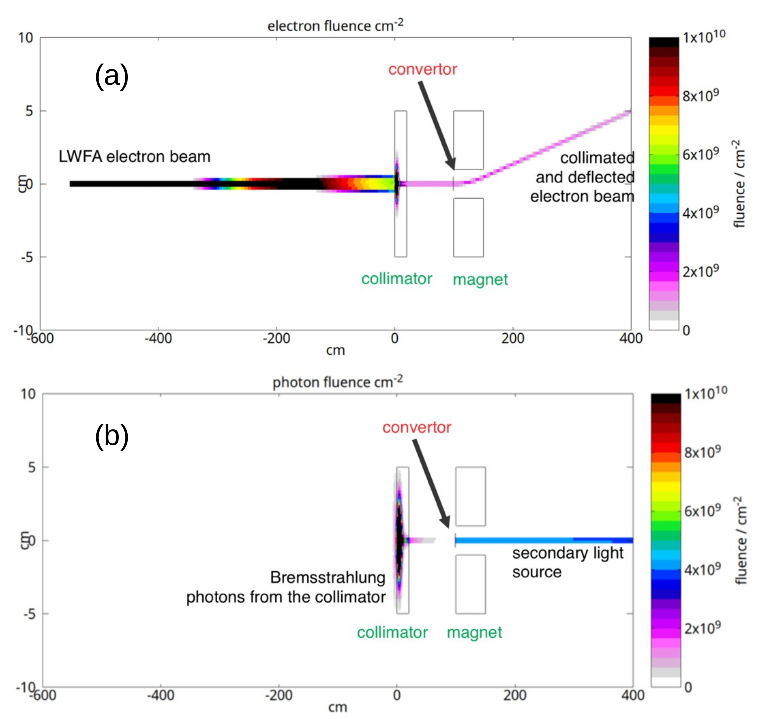}
\caption{
FLUKA CERN Monte Carlo model of the low divergence LWFA beam
 delivery to the user sample at ELI ELBA endstation for 20 cm tungsten
 collimator and 1nC 5GeV electron beam: (a) electron beam transport;
 (b) photon fluence field for 0.1.
 }
\label{ELI-figure01.fig}
\end{figure}
%%%%%%%%%%%%%%%%%%%%%%%%%%%%%%%%%%%%

While both technologies improve the quality of the beam during the
generation process, there are possibilities for further collimation and
focusing of the beam.
Introducing a 20 cm long, 0.1 mm tungsten slit improves beam divergence
to a few hundred $\mu$rad, with a quasi-flat-top profile.
However, this comes at the cost of clipping the total beam charge and
increasing secondary photon generation, as demonstrated by the Monte
Carlo FLUKA CERN model  \cite{ELI-Ref16,ELI-Ref17}
in Fig. \ref{ELI-figure01.fig}.

Conversely, beam focusing does not have these disadvantages, but
typically requires a series of bulky, heavy magnetic focusing elements.
However, Active Plasma Lens (APL) technology \cite{ELI-Ref18}
has reached the prototyping stage and is ready for deployment at a
user facility.
The APL is based on generating magnetic fields by pulsed
coaxial-to-electron-beam gas discharge current.
This compact magnetic field element (a few cms in size) enables electron
beam focusing with low astigmatism.
The current status and potential further improvements to the ELBA
Beamline electron beam parameters are summarised in Table \ref{Table.01}.

%%%%%%%%%%%%%%% Table for positrons
\begin{table}[h]
\caption{
ELBA Beamline electron beam parameters with and without beam
quality improvement technology.
Listed are: beam radius in air, $\rho$;
beam divergence, $\sigma_{\theta}$;
beam charge, $Q$;
beam energy spread at FWHM, $\Delta\E$.
}
\begin{ruledtabular}
%\begin{tabular}{p{4.5cm}p{0.5cm}p{1.5cm}p{1.5cm}p{1.5cm}p{2.cm}p{2.5cm}}
\begin{tabular}{lcccccl}
 Technology   used             &
  $\rho$                       &
 $\sigma_{\theta}$             &
 $Q$                           &
 $\Delta\E$ &
Source
\\
             &
  (mm)       &
 (mrad)      &
 (nC)        &
  &
\\
\hline
No additional technology & 6 & 3  & $>1$    & 0.1-2 GeV &Current work
\\
Guided beam (localised ionisation injection) & 2 & 1  &$\sim10^{-2}$& 10-20 \%   & Ref. \cite{ELI-Ref14}
\\
Guided beam (shock \& nanoparticle injection) & 2 & 1  &$>10^{-1}$& 10-50 \%   & Ref. \cite{ELI-Ref15}
\\
Collimation slit& 0.1 & $<0.3$ &$10^{-1}$& 10-50 \%   & Current work
\\
APL             &     & $<0.1$ &$<10^{-2}$& 10-20 \%   & Estimation   \\
\end{tabular}
\end{ruledtabular}
\label{Table.01}
\end{table}

Electron beams generated with ELBA Beamlines can be used to produce
secondary positron beams in a high-Z (e.g. tungsten) converter.
The parameters of the resulting positrons will depend on electron beam
properties, material, and dimensions of the converter.
Due to momentum conservation in thin high-Z material, positrons will
obviously be produced with parameters similar to the parent beam but
with low yield.
A thick converter will improve yield but will result in significant
scattering and variation in the parameters of positrons.
Introducing such a positron converter can be considered a future
development of the ELBA beamline for channeling experiments.

%%%%%%%%%%%%%%%%%%%%%%%%%%%%%%%%%%%%%%%%%%%%%%%%%%%%
\section{Simulation methodology  \label{Methodology}}

To simulate the propagation of ultra-relativistic electrons and
positrons in a crystalline medium we used the
commercially available scientific software
\textsc{MBN Explorer} \cite{MBNExplorer_2012}, which utilises
the relativistic classical molecular dynamics (Rel-MD)
framework \cite{MBN_ChannelingPaper_2013}.

The implemented algorithms enable the modelling of the passage
of particles over macroscopic distances with atomistic accuracy,
accounting for the interaction of a projectile with all the atoms
in the crystalline environment.

Particle trajectories are generated by accounting for randomness.
This randomness is due to:
(i) the sampling of incoming projectiles;
(ii) the displacement of lattice atoms due to thermal vibrations;
and
(iii) the probabilistic nature of the ionising collisions of the
projectile with crystal atoms.
These are incorporated into classical Rel-MD in accordance with
their probabilities
\cite{SushkoKorolSolovyov:NIMB_v569_165911_2025}.
Thus, each simulated trajectory corresponds to unique
initial conditions and specific crystalline environment, ensuring that
all trajectories are statistically independent.

An overview of the relevant algorithms and the results of
all-atom Rel-MD simulations of channeling
and radiation processes in oriented crystals
is given in
\cite{KorolSushkoSolovyov:EPJD_v75_p107_2021,CLS-book_2022}.

The simulated trajectories are used to analyse the channelling
efficiency and to characterise the emitted radiation.
The latter includes numerical calculation of spectral
distribution of radiation, $\d E(\theta_0)/\d(\hbar\om)$,
emitted by a beam of particles
and the corresponding number of emitted photons.

For each trajectory, the spectral distribution of radiation
$\d E_j(\theta_0)/\d(\hbar\om)$
($j=1,\dots, N_0$ where $N_0$ is the number of trajectories)
emitted within the cone with an opening angle of $\theta_0$
can be calculated either within the framework of classical
electrodynamics \cite{Jackson} or using the quasi-classical
formalism \cite{Baier}.
The former is adequate in the soft-photon limit, when the photon
energy, $\hbar\om$, is small compared to the energy
of a projectile particle, $\E$.
The latter approach, which accounts for the quantum recoil,
allows for an accurate quantitative description over a broader
range of the photon energies.
The \textsc{MBN Explorer} package provides options for the numerical
calculation of $\d E_j(\theta_0)/\d(\hbar\om)$ within the
framework of both approaches.
The spectral distribution of radiation
emitted by a beam of particles is calculated by averaging
the individual spectra.

The number of photons $N_{\rm ph}$ emitted
per unit time within a bandwidth (BW) $\Delta \om /\om$ is related
to the spectral distribution:
\begin{eqnarray}
N_{\rm ph}
=
{\d E(\theta_0) \over \d(\hbar\om)}
{\Delta \om \over \om}
{I \over e}
=
6.25\times 10^{6}
{\d E(\theta_0) \over \d(\hbar\om)}
{\Delta \om \over \om}
I \mbox{[nA]},
\label{eq.01}
\end{eqnarray}
where $e$ is the elementary charge,
$I$ stands for the average beam current.
The numerical factor on the right-hand side
corresponds to the current measured in nanoamperes.

At the ELBA beamline of ELI ERIC Beamlines, the electron beam is
produced in short bunches of $\Delta t = 30$ fs with a repetition
rate of 1-3.3 Hz.
The bunch charge varies according to the acceleration technique and is
typically limited to 1 nC \cite{ELI-Ref14}.
Recent measurements using the EBDS technique \cite{ELI-Ref08},
integrated charge detector and water dose deposition successfully
reached a charge density of 10 nC/cm$^2$ at 1 nC level bunch charge.
This performance can, however, potentially be improved further by
implementing a nanoparticle injection technique \cite{ELI-Ref15}.
These values result in the average and peak currents of
$I = 1-3.3$ nA and
$I_{\rm peak} \approx 33\, \mbox{kA} = 33\times10^{12}\, \mbox{nA}$,
respectively.
The number of photons per bunch is calculated using $I_{\rm peak}$ on
the right-hand side of Eq. (\ref{eq.01}).

%%%%%%%%%%%%%%%%%%%%%%%%%%%%%%%%%%%%%%%%%%%%%%%%%%%%%%%%%%%%%%%%%
\section{Case studies  \label{CaseStudies}}

\begin{figure} [h]
\centering
\includegraphics[clip,width=\textwidth]{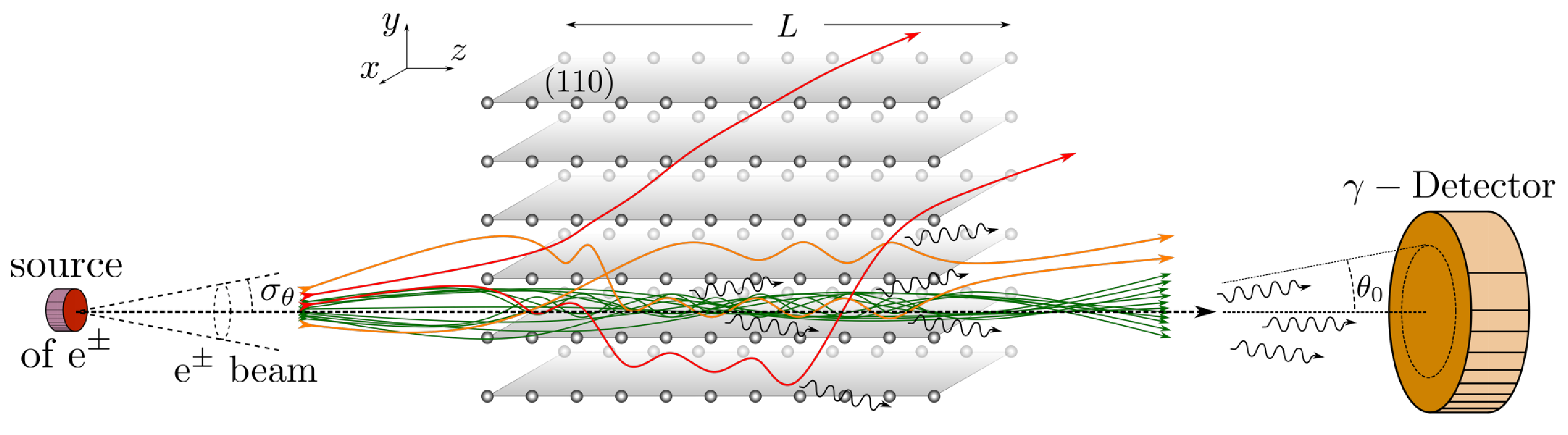}
\caption{
A schematic representation of the beam-crystal geometry.
In the reference frame $(x, y, z)$, the $y$-axis is aligned with
the [110] crystallographic axis, and the $x$- and $z$-axes lie
within the
(110) crystallographic plane.
The centerline of the incident beam is aligned with the $z$-axis.
The beam divergence $\sigma_{\theta}$ is the same in
both the $y$ and $x$ directions.
The crystal thickness along the beam's centerline is $L$.
The solid lines illustrate the particle trajectories and the short
wavy lines show the emitted radiation.
The detector collects the radiation within the cone
along the incident beam direction; the half-angle of the cone is
denoted by $\theta_0$.
}
\label{scheme.fig}
\end{figure}

Simulations were performed for $\E=1$ and 4 GeV electron beams and
for
a $\E=4$ GeV positron beam incident on oriented silicon single crystal
along the (110) planes.
Figure \ref{scheme.fig} illustrates the beam-crystal
geometry.
% \footnote{\textcolor{red}{Figure to be re-drawn in a more artistic/attractive style.}}.
The $y$-axis is orthogonal to the $(110)$ crystallographic plane.
The thickness of the crystals ($L$) along the centerline of
the incident beam (the $z$-axis) was fixed at 0.2 mm for electrons
and was varied from 0.2 to 2 mm for positrons.

%%%%%%%%%%%%%%%%%% Beam-Crystal geometry
%%%%%%%%%%%%%%%%%%%%%%%%%%%%%%%%%%%%

%
Initial coordinates $x_0$ and $y_0$ were generated using uniform
distributions at the crystal entrance.
% \textcolor{red}{Since motion in the $x$ direction does not
% affect the dynamics of the
% channelling oscillations, which occur in the $y$ direction, the angular
% divergence of the beam in the $x$ direction was disregarded.
% \footnote{\textcolor{red}{To be checked! The dynamics is not affected but the
% intensity of radiation within a finite cone is affected!}}
% }%
To generate the initial entrance angles $\theta_{x,y}= v_{x,y}(0)/c$
(here, $v_{x,y}(0)$ are the velocities at the entrance and $c$ is
the speed of light)  of an
incident particle in both directions, a normal distribution with
a standard deviation of $\sigma_{\theta}$ was used.
For each energy $\E$ the simulations were performed for
'wide', 'narrow' and 'nearly collimated' beams.
The corresponding values of $\sigma_{\theta}$ were chosen to be equal to
$5\theta_L$, $\theta_L/2$ and  $\theta_L/5$, respectively.
Here, $\theta_L=\sqrt{2U_0/\E}$ is Lindhard's critical angle
\cite{Lindhard}, with $U_0$ representing the depth of the interplanar
potential.
For the (110) planar channel in silicon $U_0\approx 20$ eV
(see, for example, \cite{KorolSushkoSolovyov:EPJD_v75_p107_2021,%
ScandaleTaratin:PhysRep_v815_p1_2019}).
This results in values $\theta_L\approx 200$ and 100 $\mu$rad for
$\E=1$ and 4 GeV, respectively.

One should stress that, for electrons, the ’wide’-beam values
align with those presently available at the ELI facility
\cite{ELI-Ref14}.
The plan for the development of 'narrower' beams, reaching divergences
of just a few hundred $\mu$rad, is currently underway for the
ELI beamline.
%
% We note that for electrons the 'wide'-beam values correspond to those
% currently available at the
% \textcolor{red}{ELI facility [Ref]}.
% \textcolor{red}{
% Beams with 'narrow' divergence are planned to be available in the
% near future, once the necessary amendments have been made to the
% experimental setup.
% }
The values $\sigma_{\theta}=\theta_L/2$, being half the
critical angle, ensure that a sufficiently large fraction of the
beam particles is accepted
in the channelling mode at the crystal entrance.
Simulations were performed for 'nearly collimated' beams to
assess whether a further decrease in divergence results in a fairly
noticeable increase in radiation intensity.

Previous numerical analyses
\cite{SushkoKorolSolovyov:NIMB_v569_165911_2025,%
KorolSolovyov:EPJD_v80_25_2026}
have shown that for electrons of energy within the sub-GeV up to ten-GeV
range, accounting for ionising collisions does not significantly affect
either the channelling efficiency or the radiation intensity.
This observation has a direct practical consequence: numerical simulations
carried out within the framework of classical Rel-MD without accounting for
ionising collisions are much faster, yet produce the same result as those
in which ionising collisions are accounted for.
In contrast, for positrons, the ionising collisions significantly reduce
the channelling efficiency and the radiation intensity.
Therefore, the results for electrons reported in this paper were
obtained without accounting for the ionising collisions whereas
 the results for positrons were obtained with ionising collisions.

% \textcolor{red}{
% During the passage of the particles, the crystalline environment was
% generated dynamically accounting for the random displacement of the atoms
% from their nodal positions due to thermal vibrations.
% }

 %%%%%%%%%%%%%%%%%%%%%%%%%%%%%%%%%%%%%%%%%%%%%%%%%%%%%%%%%%%%%%%%%
\subsection{Case study I: 1 \& 4 GeV electrons  \label{CaseStudy_I}}

Collisions between a channelling particle and crystal atoms can lead to
an increase in the particle's transverse energy.
This results in a gradual decrease in the number of particles
accepted in the channelling mode at the entrance with penetration
distance $z$.
This dependence can be characterised in terms of the ratio
$f_{\rm ch,0}(z) = N_{\rm ch,0}(z)/N_0$, where $N_0$ is the
total number of incoming particles and $N_{\rm ch,0}(z)$ is the
number of particles that have been accepted and channeled up to $z$.
Collisions can also result in rechannelling events, whereby
non-channelling particles lose transverse energy and become
captured in the channelling mode at some point within a crystalline
medium.
Taking rechannelling into account, the number $N_{\rm ch} (z)$
of channelling particles at distance $z$ exceeds $N_{\rm ch,0}(z)$.

The aforementioned fractions can be used to quantify the
dechanneling and rechanneling effects by introducing two mean
distances \cite{KorolSushkoSolovyov:EPJD_v75_p107_2021}:
(i) the penetration length $L_0$, which is
the mean length of the initial channeling segment of the trajectory of
an accepted particle, and (ii) $L_{\rm ch}$, which is the mean sum
of all channeling segments in a particle's trajectory:
\begin{eqnarray}
L_0 = \calA^{-1}\int_0^L f_{\rm ch,0}(z) \d z\,,
\qquad
L_{\rm ch} = \int_0^L f_{\rm ch}(z) \d z\,.
\label{CaseStudies:eq.01}
\end{eqnarray}
Here, the acceptance factor, $\calA = f_{\rm ch,0}(0)$,
is equal to the fraction of the particles accepted at the entrance.
Note that $L_0$ is calculated by taking into account only accepted particles,
whereas $L_{\rm ch}$ is the mean value per trajectory,
calculated using the
total number $N_0$  of trajectories.

\begin{figure} [h]
\centering
\includegraphics[clip,width=\textwidth]{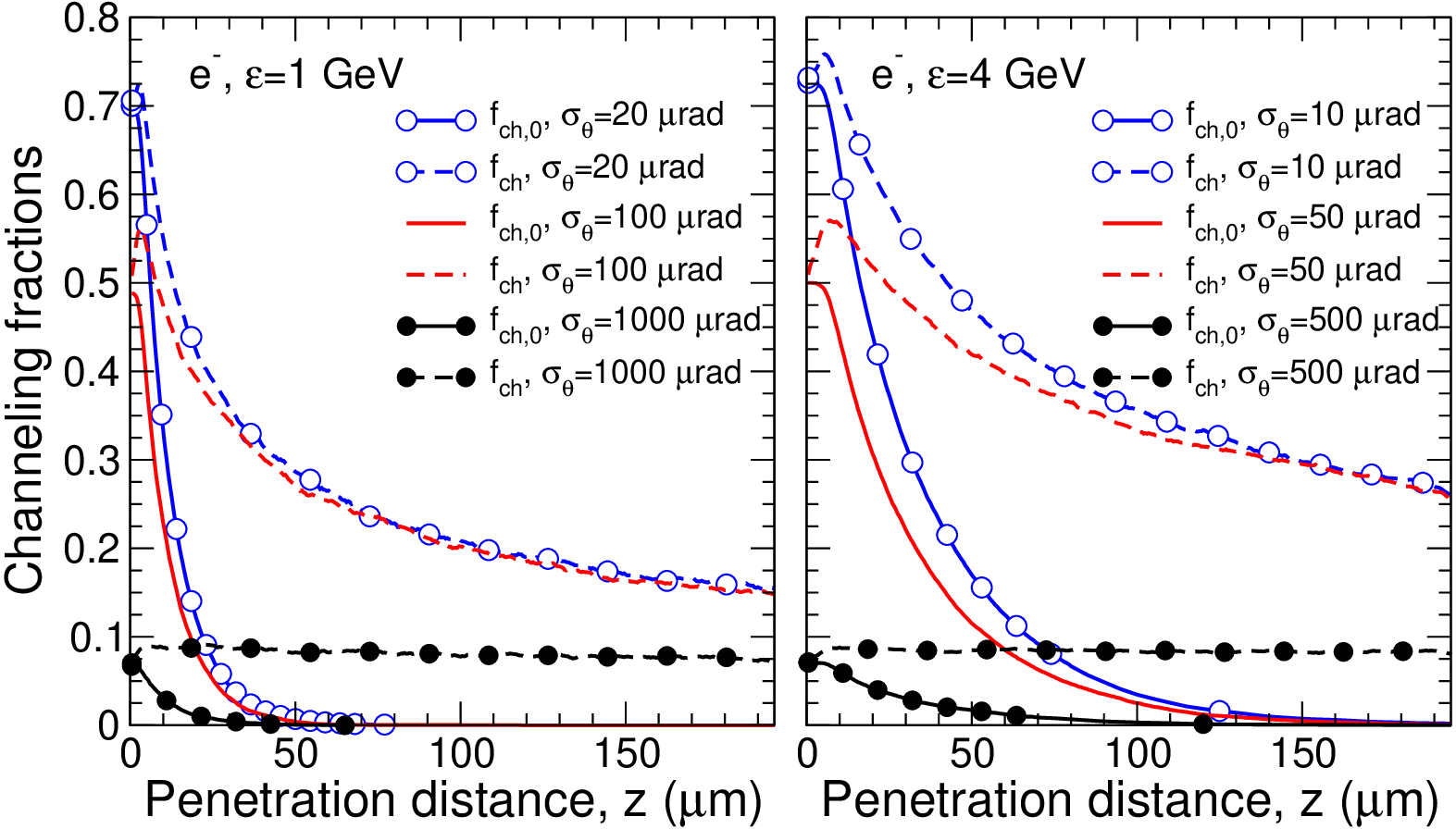}
\caption{
Channeling fractions $f_{{\rm ch}, 0}(z)$ (solid lines) and
$f_{\rm ch}(z)$ (dashed lines) versus penetration distance $z$
for 1 GeV (left panel) and 4 GeV (right panel) electrons in
$L=200$ $\mu$m thick oriented silicon (110) crystal.
The dependencies presented in each graph refer to three values
of the RMS beam divergence $\sigma_{\theta}$, as indicated in the legends.
For each energy, the $\sigma_{\theta}$ values correspond (in the ascending
order) to $\theta_L/5$ (open circles),
$\theta_L/2$ (no symbols) and  $5\theta_L$ (filled circles).
See also explanations in the text.
% with
% $\theta_L\approx 200$ and 100 $\mu$rad being Lindhard's critical angle
% for 1 and 4 GeV electrons, respectively.
}
\label{e-Fraction.fig}
\end{figure}

Figure \ref{e-Fraction.fig} shows the results of the numerical analysis
of the channelling efficiency for $1$ GeV (left panel) and 4 GeV (right
panel) electrons in Si(110).
The dependencies shown refer to the fractions $f_{\rm ch,0}(z)$ (solid
lines) and $f_{\rm ch}(z) = N_{\rm ch}(z)/N_0$ (dashed lines), which
were calculated for 'wide' (solid circles), 'narrow'
(no symbols) and 'nearly collimated' (open circles) beams.
For each energy, the corresponding values of the
RMS beam divergence, $\sigma_{\theta}$, are indicated in the legends.
Approximately $N_0=2\times10^4$ trajectories were simulated and analysed
for each case study.

% %%%%%% Fractions, e- 1 & 4 GeV
%%%%%%%%%%%%%%%%%%%%%%%%%%%%%%%%%%%%

In the case of the 'wide' beam, whose RMS divergence is five times
larger than the critical angle, only a small fraction
(less than ten per cent) of particles are accepted at the entrance.
Due to rechannelling, the number of channeling particles is approximately
the same at each distance $z\leq L=200$ $\mu$m.
It can be seen that $f_{\rm ch}(z) \approx \calA$ for both values of
the beam energy.

When comparing the dependencies for the 'narrow' and
'nearly collimated' beams,
one can see that the fractions for the latter are notably higher
at smaller distances $z \lesssim L_0 \ll L$ due to a larger number of
accepted particles, see Table \ref{Table.02}.
For larger distances, where the $f_{\rm ch,0}(z)$ fractions drop off,
the $f_{\rm ch}(z)$ fractions are very close for both beams.
Therefore, the number of channelling particles is approximately the same
in most of the crystal.
Consequently, one can expect the intensity of the channelling radiation
emitted by the channelling particles to be similar for both beams.

Table \ref{Table.02} summarises the values of $\calA$,
$L_0$ and $L_{\rm ch}$ calculated for all cases.
The statistical uncertainties indicated in the table are due to the
finite number of the simulated trajectories.
For each energy, the penetration length $L_0$
 remains constant across all beam divergence values.
The total length of the channeling segments $L_{\rm ch}$
for the 'narrow' and 'nearly collimated' beams is close and greatly
exceeds that for a 'wide' beam.

%%%%%%%%%%%%%%% Table for electrons
%\begin{turnpage}
\begin{table}[h]
\caption{Acceptance $\calA$,
penetration length $L_0$
and total length of channelling segments $L_{\rm ch}$,
Eq. (\ref{CaseStudies:eq.01}),
for 1 and 4 GeV electrons incident on Si(110).
Values of $L_0$ and $L_{\rm ch}$ are in microns,
values of $\sigma_{\theta}$ in $\mu$rad.
}
\begin{ruledtabular}
\begin{tabular}{cccccccc}
      &\multicolumn{3}{c}{$\E=1$ GeV}
    & &\multicolumn{3}{c}{$\E=4$ GeV}\\
$\sigma_{\theta}$ & 20          & 100          &   1000   &
                  & 10          &  50          &    500   \\
\hline
$\calA$           & 0.70        &  0.49        &    0.067 &
                  & 0.73        &  0.50        &    0.071      \\
\hline
$L_0$             & $12.7\pm0.3$& $12.5\pm0.4$ &$12.6\pm0.9$&
                  & $36.0\pm0.8$& $36.2\pm1.0$ &$35.7\pm2.7$          \\
$L_{\rm ch}$      & $50.9\pm0.9$& $45.5\pm0.8$ &$16.2\pm0.6$&
                  & $80.8\pm1.1$& $73.1\pm1.2$ &$16.8\pm0.9$         \\
\end{tabular}
\end{ruledtabular}
\label{Table.02}
\end{table}
%\end{turnpage}

\begin{figure} [h]
\centering
\includegraphics[clip,width=0.49\textwidth]{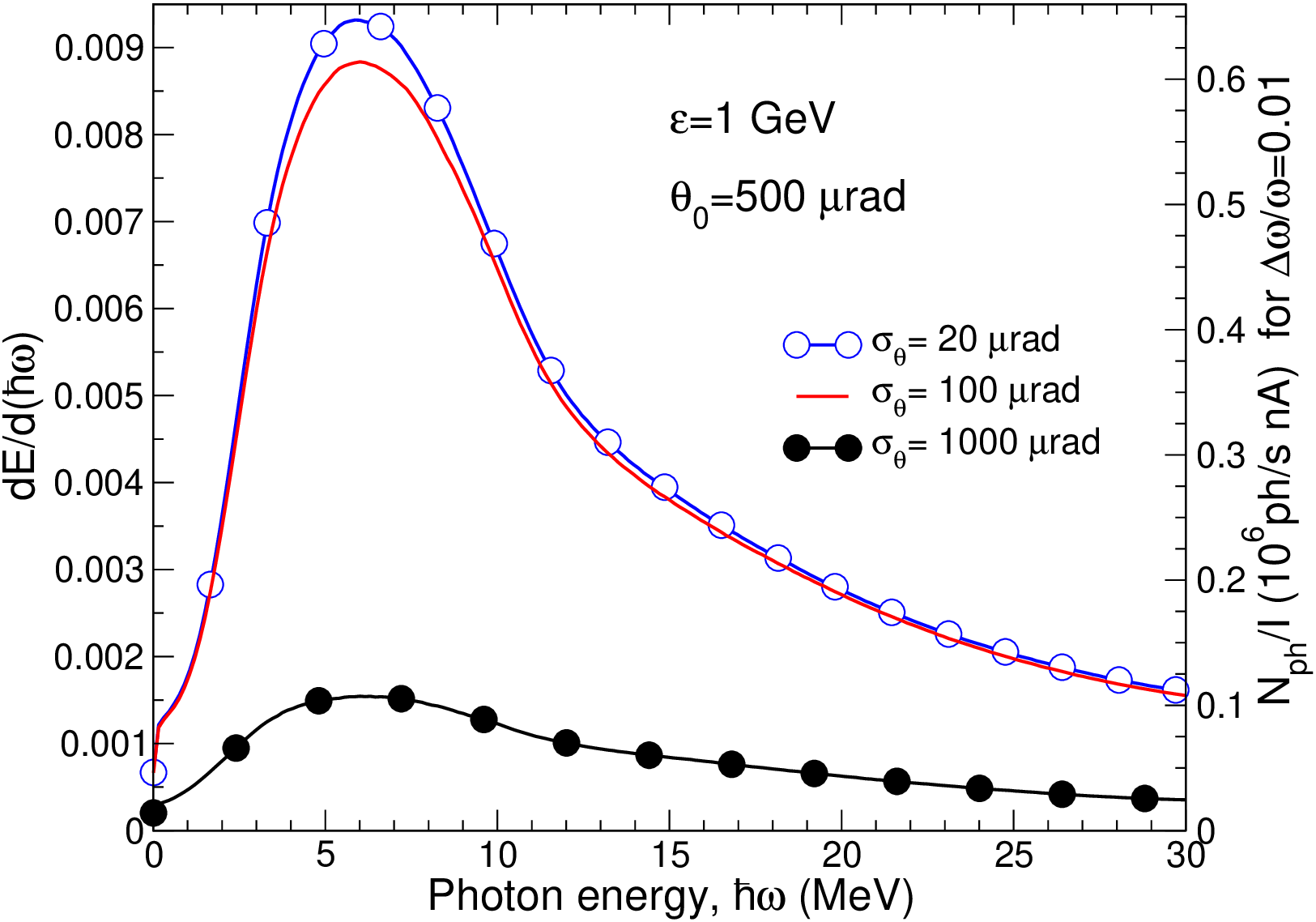}\hspace*{0.2cm}
\includegraphics[clip,width=0.49\textwidth]{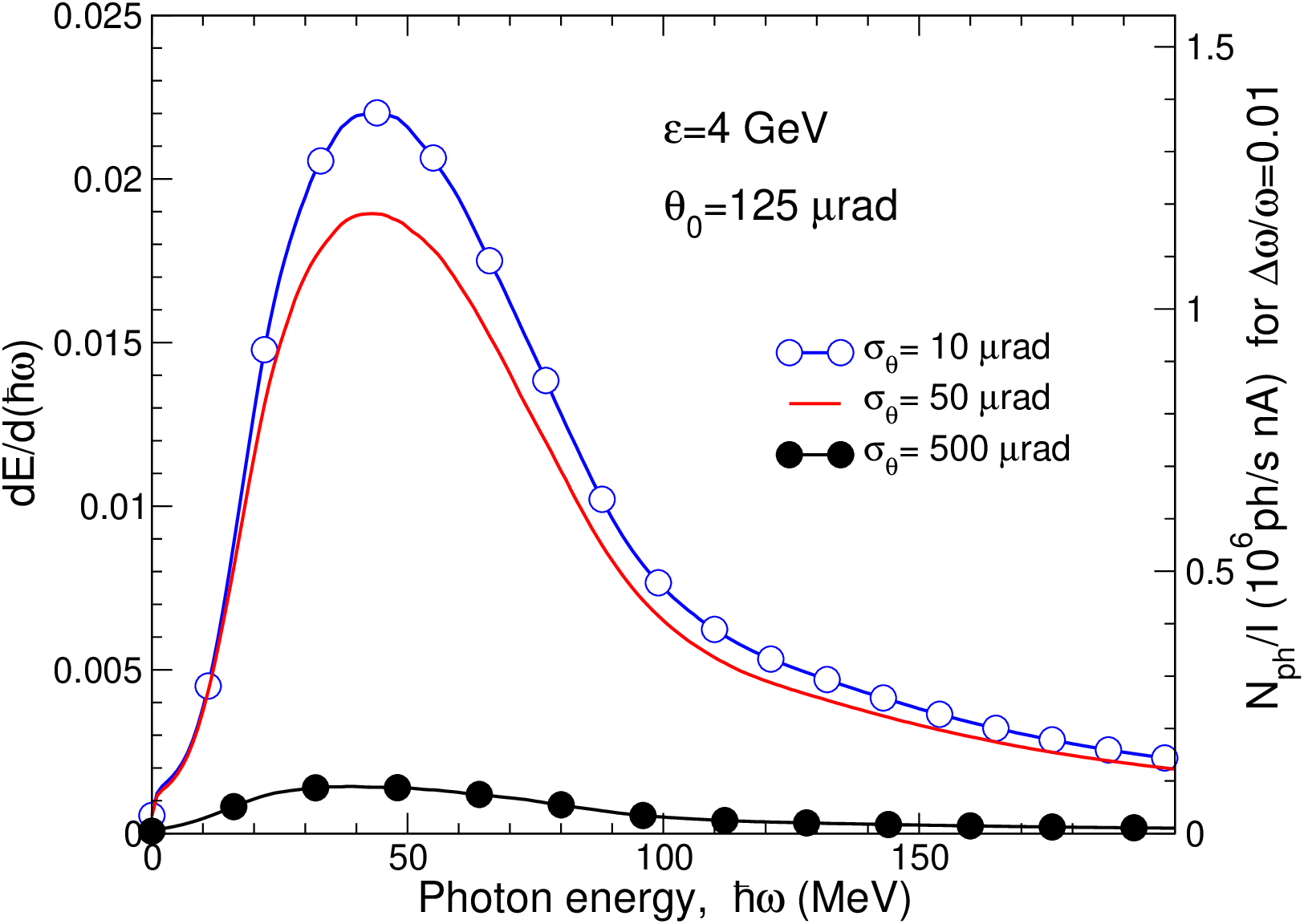}
\caption{
Spectral distribution of radiation (left axes)
and number of photons per second and per 1 nA of the beam current
(right axes) emitted by 1 GeV (left graph) and 4 GeV (right graph)
electrons incident along the (110) planar direction
in silicon crystal of thickness 200 $\mu$m.
The curves with open circles (blue), without symbols (red)
and with filled circles (black)
correspond to the 'nearly collimated', 'narrow' and 'wide' beams,
respectively.
The corresponding beam divergences are indicated in the legends.
The graphs refer to the emission cone $\theta_0\approx 1/\gamma$.
The number of photons is calculated for the BW
$\Delta \om /\om = 0.01$.
For both energies, the value of $\d E/\d(\hbar\om)$
of the bremsstrahlung radiation  in the amorphous
medium (not shown in the figure)
is approximately $7.5 \times 10^{-4}$ in the whole range of
photon energies.
}
\label{e-Spectra-1.fig}
\end{figure}

The discrepancies and similarities in the dynamics of dechanneling and
rechanneling mentioned above for beams with different divergences are also
evident in the photon emission spectra presented in
Fig. \ref{e-Spectra-1.fig}.
The two graphs present the results of calculations of the spectral
distributions of radiation (left vertical axes) and of
$N_{\rm ph}/I$ (right vertical axes)
performed for 1 and 4 GeV
'nearly collimated', 'narrow' and 'wide' beams electron beams.
The emission cone considered is $\theta_0\approx 1/\gamma$.
The corresponding values in microradians are indicated in each graph.

%
% %%%%%% Spectra, e- 1 & 4 GeV
%%%%%%%%%%%%%%%%%%%%%%%%%%%%%%%%%%%%

For both beam energies, the spectral distributions are dominated
by the peaks of channelling radiation.
Even for the 'wide' beam, the peak intensity greatly exceeds that
of the incoherent bremsstrahlung radiation emitted by the projectiles
in 200 $\mu$m thick amorphous silicon
(the value of $\d E/\d(\hbar\om)$ is indicated in the caption).
The peak values of the 'nearly collimated' beam exceed those of the 'narrow' beam by only about 5–10 per cent.
This is in accordance with the difference in the $L_{\rm ch}$ values,
see Table \ref{Table.02}.
The peak intensity of the 'wide' beam is approximately 6 and 14 times
lower for electrons with energies of 1 and 4 GeV, respectively.
This significant decrease in intensity is due to two main reasons.
Firstly, the divergence of the 'wide' beam in the $y$ direction is
5 times larger than Lindhard's critical value, which considerably
reduces the  number of channeling particles that emit
channeling radiation.
Secondly, the beam divergence in the $x$ direction exceeds
the values of $\theta_0=1/\gamma$ by factors 2 and 4 for
1 and 4 GeV electrons, respectively.
Consequently, only some of the channeling radiation emitted by
the particles entering the crystal at angles $\theta_x> \theta_0$
is collected within the cone along the incident beam direction.
The ratio $\sigma_{\theta}/\theta_0$ is larger for the
the 4 GeV beam, therefore, this effect is more pronounced in this case.

Note that the $N_{\rm ph}/I$ values shown correspond to an average
current $I$, of the order of magnitude of 1 nanoampere, depending on the
selected laser repetition rate.
This aligns with the average current at ELI.
Given a sub-mrad beam divergence, this enables a current density in
the range of nA/cm$^2$ at the vacuum-to-air interface plane.

With the electron beams currently available at the ELI facility,
we can state that the number of photons emitted within a narrow cone and
bandwidth of 0.01 can be on the level of $10^5$ ph/s
\footnote{Note that in many operational or developing Compton
gamma-ray LSs the $\Delta \om /\om$ value is an order
of magnitude higher (see, for example, Table I in
Ref. \cite{KorolSushkoSolovyov:PRAB_v27_p100703_2024}).
Therefore, the $N_{\rm ph}/I$ values achievable in the CLS considered
in the present paper can be increased correspondingly.}.
A decrease in the divergence by a factor of 5 would
increase the $N_{\rm ph}/I$ values by six to fourteen times.

With reference to the 'narrow' beam specifically, two other factors that
lead to an increase in the photon yield should be mentioned.
Firstly, as it is seen in Fig. \ref{e-Fraction.fig}, the fraction of
particles moving in the channeling mode at $z=200$ $\mu$m is still
fairly large.
This suggests that thicker crystals can be used to increase
$N_{\rm ph}/I$.
Secondly, collecting the radiation within a larger emission cone has a
similar effect.
The data in Fig. \ref{e-Spectra-1.fig} refer to the emission cone
$\theta_0\approx 1/\gamma$, within which an ultra-relativistic particle
emits a noticeable fraction of the radiation, but not all of it.
Therefore, the enhancement of the radiation yield can be achieved by
increasing the emission cone.

\begin{figure} [h]
\centering
\includegraphics[clip,width=0.49\textwidth]{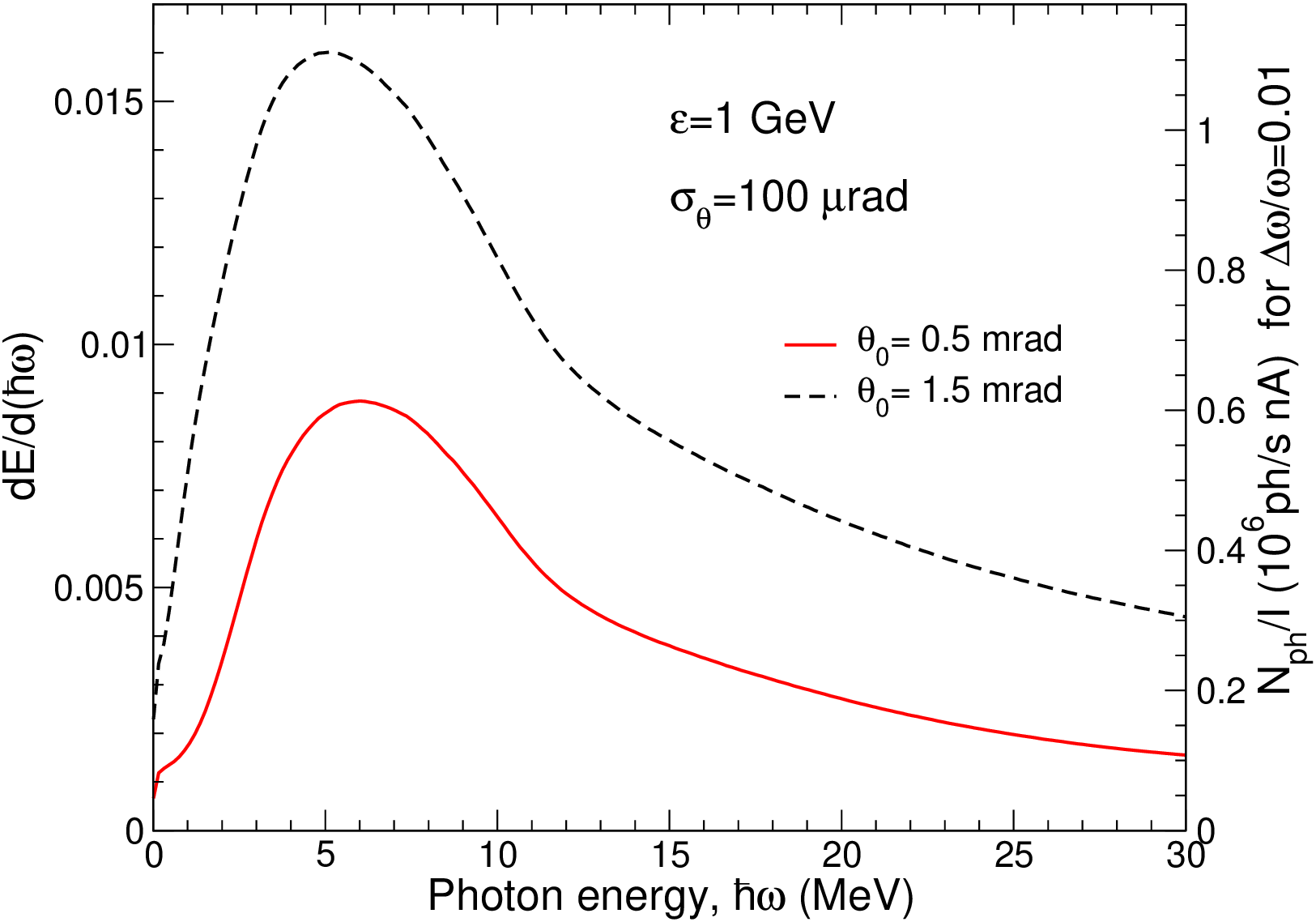}\hspace*{0.2cm}
\includegraphics[clip,width=0.49\textwidth]{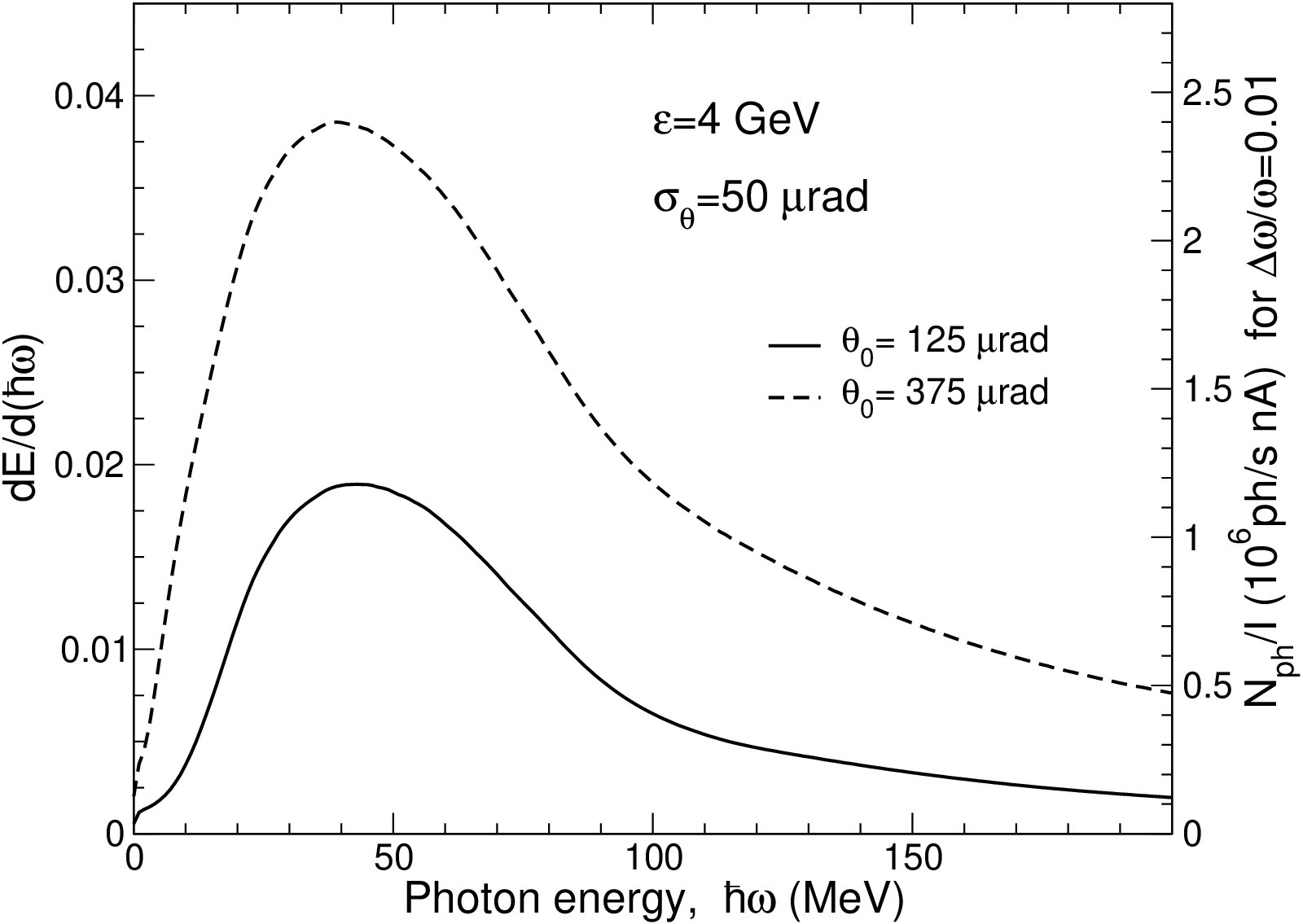}
\caption{
Spectral distribution of radiation (left axis)
and number of photons per second and per 1 nA of the beam current
(right axis) emitted by 1 (left panel) and 4 (right panel)
GeV electrons within the cones of
$\theta_0 \approx 1/\gamma$ (solid line)
and
$\theta_0 \approx 3/\gamma$ (dashed line),
respectively.
The beam divergences $\sigma_{\theta}$ indicated correspond
to the 'narrow' beams.
The number of photons is calculated for the bandwidth
$\Delta \om /\om = 0.01$.
}
\label{e-Spectra-2.fig}
\end{figure}

Figure \ref{e-Spectra-2.fig} compares the spectral distributions of
radiation emitted by the 'narrow'
($\theta_L/2=50$ $\mu$rad) 1 and 4 GeV electron beams
within the emission cones of approximately
$1/\gamma$ (solid lines)
and $3/\gamma$ (dashed lines).
The latter value corresponds to the cone that collects virtually all the
emitted radiation.
For both energies,
the peak values are two times larger for the wider cone.

% %%%%%% Spectra, e- , 1&4 GeV, different theta_0
%%%%%%%%%%%%%%%%%%%%%%%%%%%%%%%%%%%%

\begin{figure} [h]
\centering
\includegraphics[clip,width=0.75\textwidth]{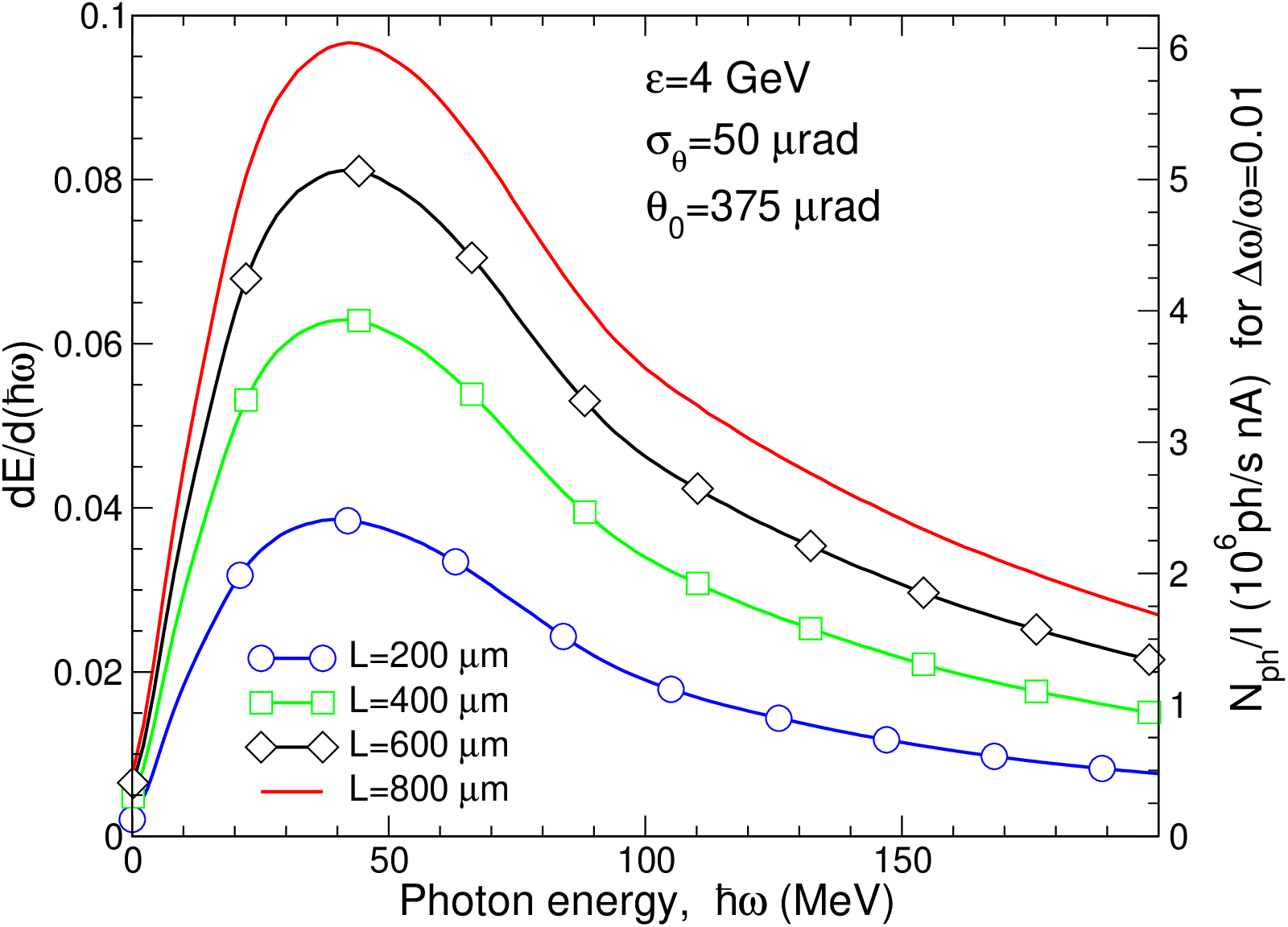}
\caption{
Spectral distribution of radiation (left axis)
and number of photons per second and per 1 nA of the beam current
(right axis) emitted within the cone $\theta_0 \approx 3/\gamma$
by 4 GeV electrons incident on the
crystals of different thickness $L$ as indicated in the caption.
The beam divergence $\sigma_{\theta}=50$ $\mu$rad corresponds
to the 'narrow' beam.
The number of photons is calculated for the bandwidth
$\Delta \om /\om = 0.01$.
}
\label{e-Spectra-3.fig}
\end{figure}

To address the aforementioned  possibility of using a thicker crystal
to increase the yield of photons, we present Fig. \ref{e-Spectra-3.fig}.
Here, the spectra calculated for the crystal thicknesses of
200, 400, 600 and 800 $\mu$m are compared for
the 'narrow' ($\sigma_{\theta}=50$ $\mu$rad), 4 GeV electron beam.
All curves refer to the  emission cone of approximately
$3/\gamma$.
% %%%%%% Spectra, e- , 1&4 GeV, different theta_0
%%%%%%%%%%%%%%%%%%%%%%%%%%%%%%%%%%%%

%%%%%%%%%%%%%%%%%%%%%%%%%%%%%%%%%%%%%%%%%%%%%%%%%%%%%%%%%%%%%%%%%
\subsection{Case study II: 4 GeV positrons  \label{CaseStudy_II}}

Construction of a high-quality LWFA driven positron beam is a future ambition,
which, in the first stage, can be implemented using
various tungsten converters inserted into the electron beams at the ELI
ELBA beamline.
Positron beams produced in the sub-GeV energy range can potentially be
collimated by the Permanent-Magnet Quadrupole (PMQ) focusing systems
\cite{ELI-Ref19} to achieve higher beam quality.

However, it is instructive to compare the yields of radiation emitted
by electron and positron beams with the same parameters.

Unlike electrons, positrons channel between the atomic planes.
This region has a low content of crystal electrons and is virtually free
of nuclei.
Consequently, positrons experience much softer collision with crystal
atoms, resulting in a much lower dechanneling rate.
For the same energy and crystal environment,
the dechanneling length of a positron exceeds that of an electron by an
order of magnitude \cite{ChannelingBook2014}.
Therefore, the thickness of a crystalline target can be increased by the
same factor, leading to a much larger values for the number
of photons
\cite{KorolSushkoSolovyov:PRAB_v27_p100703_2024}.

% %%%%%% Spectra, e+ , 4 GeV, different L
\begin{figure} [h]
\centering
\includegraphics[clip,width=0.49\textwidth]{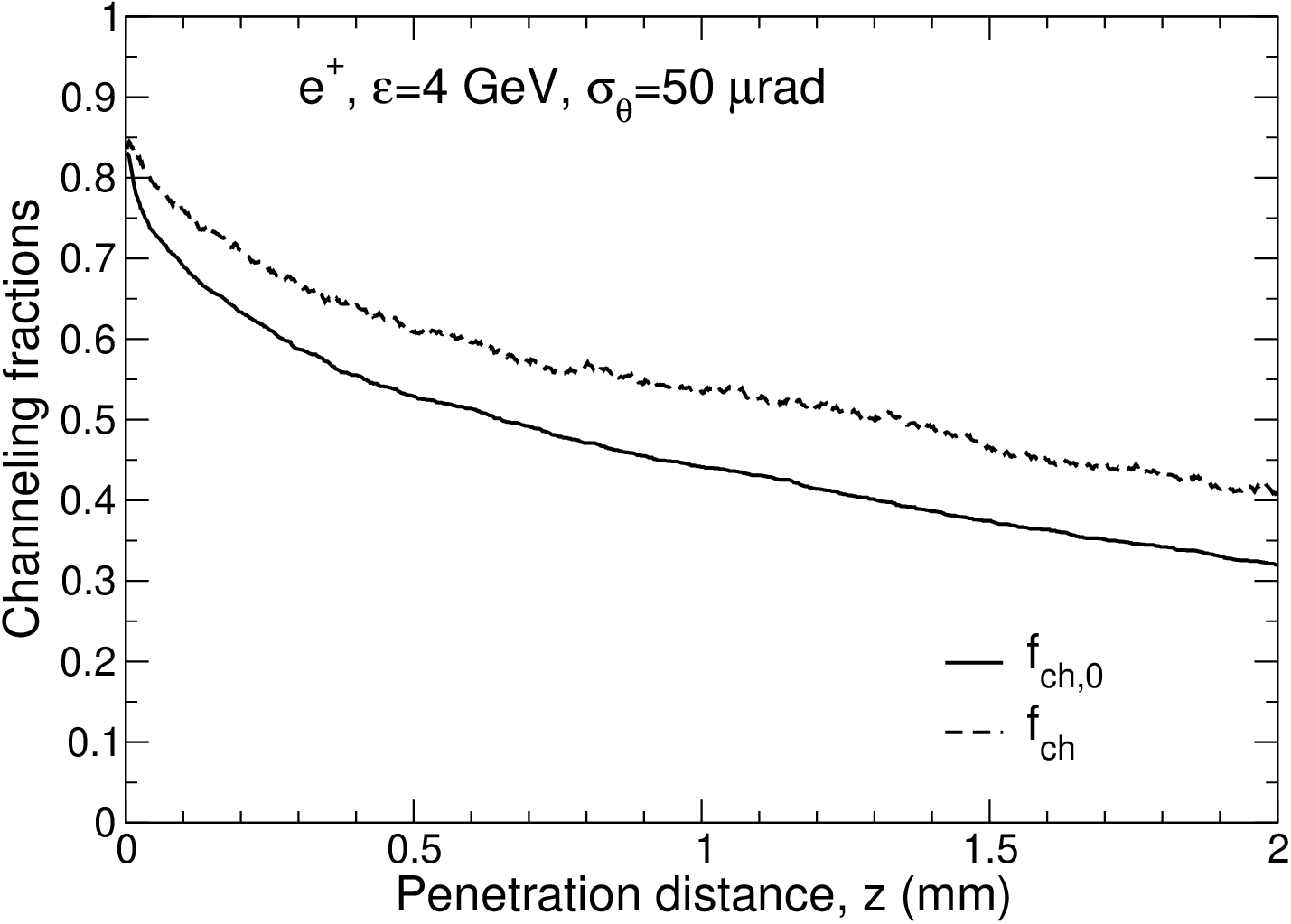}\hspace*{0.2cm}
\includegraphics[clip,width=0.49\textwidth]{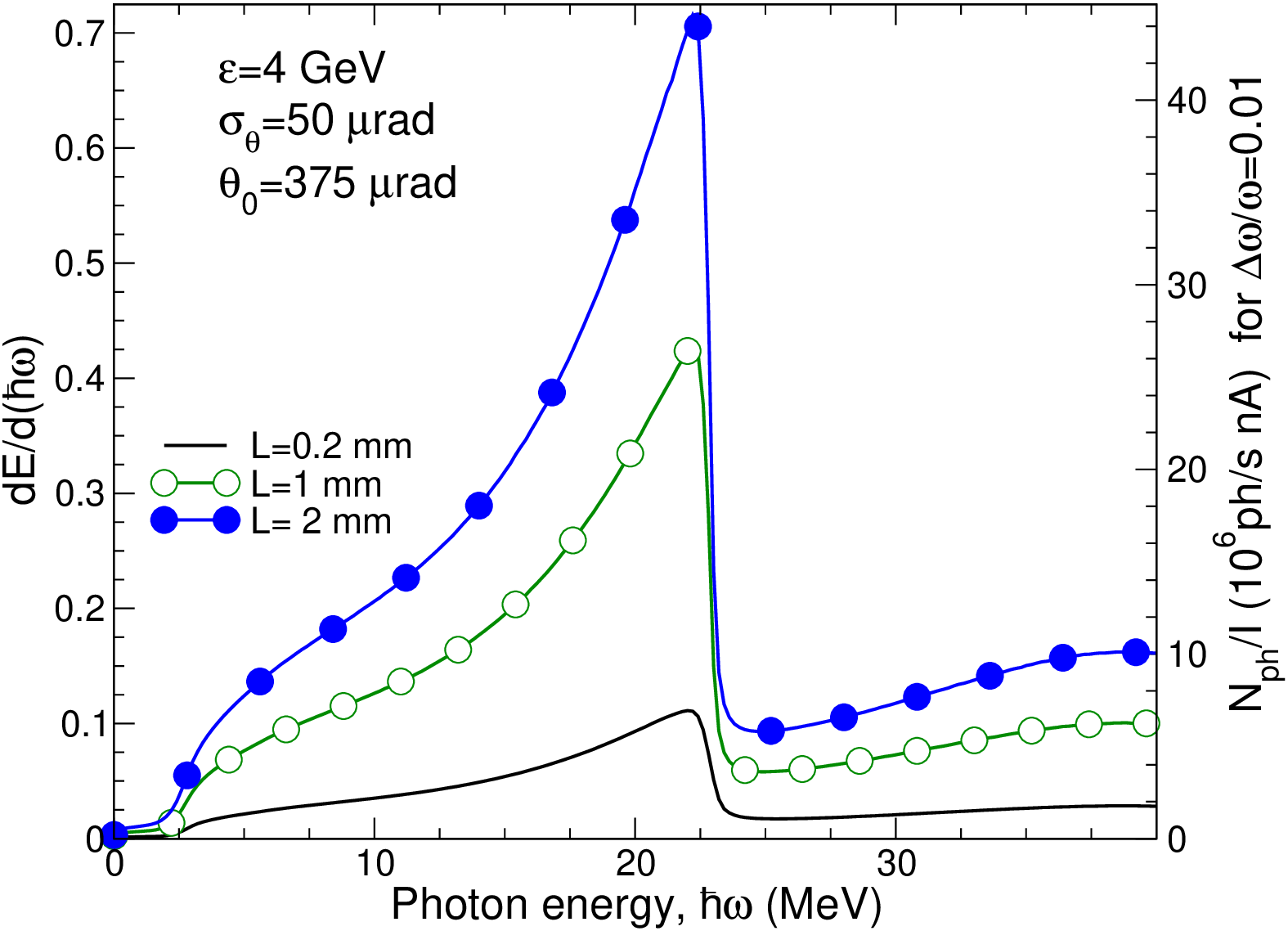}
\caption{
\textit{Left:}
Channeling fractions $f_{\rm ch, 0}(z)$ (solid lines) and
$f_{\rm ch}(z)$ (dashed lines) versus penetration distance for 4 GeV
positrons in $L=2$ mm thick oriented Si(110) crystal.
\textit{Right:}
Spectral distribution of radiation (left axis)
and number of photons per second and per 1 nA of the beam current
(right axis) emitted by 4 GeV positrons in Si(110) crystals
of thicknesses $L= 0.2$ mm (no symbols), 1 mm (open circles) and 2 mm
(filled circles) within the cone of 375 $\mu$rad.
The number of photons is calculated for the bandwidth
$\Delta \om /\om = 0.01$.
The dependencies shown in both graphs refer to the beam divergence
of 50 $\mu$rad.
}
\label{p-Spectra.fig}
\end{figure}
%%%%%%%%%%%%%%%%%%%%%%%%%%%%%%%%%%%%

Figure \ref{p-Spectra.fig}\textit{left} presents the evolution of the
channeling fractions, $f_{\rm ch, 0}(z)$ and $f_{\rm ch}(z)$, of 4 GeV
positrons propagating along the (110) planar direction in 2 mm
thick silicon  crystal.
These dependencies refer to the 'narrow' beam divergence
of $\sigma_{\theta} =50$ $\mu$rad.
Comparing these dependencies to corresponding fractions
for electrons (see the graph on the right in Fig. \ref{e-Spectra-1.fig})
clearly shows that the channeling efficiency of the positrons
is much higher.

Figure \ref{p-Spectra.fig} compares the spectral distributions of
radiation emitted by the positrons.
The calculations were performed for the crystal thicknesses of
$L=0.2$, 1 and 2 millimeters.
All curves refer to a cone with an opening angle  of
$375\ \mu \mbox{rad} \approx 3/\gamma$, which collects nearly
all radiation emitted by the particles.

Comparing the shape of the spectral distributions for positrons
and electrons, it can be seen that the radiation by positrons is
distributed over a much narrower range of photon energies.
This is due to the almost harmonic nature of positron channeling
oscillations.
Consequently, the spectral distribution of the channeling radiation
emitted by positrons closely resembles that of undulator radiation,
where an increase in radiation intensity occurs near equally spaced
harmonics (see, for example, Ref. \cite{Barbini_EtAl_1990}).
Electron channeling oscillations are strongly anharmonic, resulting
in a broader distribution of the channeling radiation.

The distributions shown in Fig. \ref{p-Spectra.fig}\textit{right}
refer to the emission in the first harmonic.
The peak value of the distributions at
$\hbar\om_{\max}\approx 22$ MeV, corresponds to the
radiation emitted in the direction of the incident beam
(i.e., on-axis radiation with an emission angle of $\theta=0$).
Off-axis radiation, with an emission angle greater than zero,
is emitted at lower energies
and its intensity gradually decreases with the emission angle.
This explains the asymmetry in the spectral distributions:
a sharp decrease right beyond $\hbar\om_{\max}$ and
a smooth decrease from the peak value in the region
$\hbar\om <\hbar\om_{\max}$.

For the purposes of this paper, it is important to emphasise that
using positron beams with the same parameters as electron beams,
% which are \textcolor{red}{either currently available at ELI or
% planned for the near future},
allows for a significant increase in
the photon yield from CLSs based on single oriental crystals.
Figure \ref{p-Spectra.fig} shows that, for a narrow bandwidth
of one per cent, the number of emitted photons in the
10–20 MeV range can be as high as
$(2-4)\times 10^7$ ph/s for an average beam current of 1 nA.
These numbers are order of magnitude larger than for the electrons.

%%%%%%%%%%%%%%%%%%%%%%%%%%%%%%%%%%%%%%%%%%%%%%%%%%%%%%%%%%%%%%%%%
\section{Conclusions  \label{Conclusions}}

Developing and implementation of electron channeling-based photon
sources requires intense, GeV-class electron beams with high
bunch charge and density.
It also requires the insertion of crystalline targets directly into the
beam path.
At conventional RF-based LINAC and storage ring facilities, these
experiments are often hindered by limited beamtime availability and
safety risks that require rather complex mitigation processes.
In contrast, the Laser Wakefield Acceleration (LWFA) technology
implemented at the ELI ELBA beamline routinely grants users with this
essential access for such experiments.
Current evaluations of the LWFA accelerator, alongside simulations of
electron channeling using parameters matching the ELI ELBA beamline,
confirm the feasibility of generating gamma beams via
LWFA-electron beam–driven CLSs.
Due to its compact footprints, a CLS system can be further developed
into a high-TRL (Technology Readiness Level) device, serving as an
alternative, high-yield, ultra-short pulse gamma source for LWFA users.
Furthermore, recently demonstrated beam-quality enhancement technologies
with potential for implementation of novel APL techniques and delivery of
secondary positron beams, indicate the possibility of reaching
performance of CLS photon source unachievable with conventional
technologies.
Finally, compared to these conventional techniques listed above, pairing
LWFA with CLS relaxes the strict electron beam requirements (such as
micro-emittance and specific beam cross-sections) that are notoriously
difficult for laser-plasma sources to achieve, while fully exploiting
their unique characteristics such as high bunch charge and time
structure.

In this paper, we demonstrate the feasibility of novel gamma-ray CLSs by exposing
single crystals to electron beams available at the ELI-ELBA facility.
To achieve this, we performed rigorous numerical simulations of the channeling
efficiency and characterisation of the emitted radiation for 1 and 4 GeV electrons
incident on single silicon crystals of different thicknesses along the (110) plane.
The simulations were performed using both the current divergence and a smaller,
more realistic value for the near future.
We demonstrate that a photon yield of $10^6-10^7$ photons/s can be achieved
in the photon energy range of 5 to 50 MeV.
A similar analysis was also performed for a 4-GeV positron beam incident on
much thicker targets, up to 2 mm oriented silicon crystal, looking further ahead.
In this case, the photon yield can be increased by an order of magnitude.

%%%%%%%%%%%%%%%%%%%%%%%%%%%%
\section*{Acknowledgements}

The work was supported by the European Commission through
% the N-LIGHT Project within the H2020-MSCA-RISE-2019 call (GA 872196)and
the Horizon  Europe EIC-Pathfinder-Project TECHNO-CLS
(Project No. 101046458) and
the HORIZON TMA MSCA Staff Exchanges (GA ID: 101299799).
We acknowledge the Frankfurt Center for Scientific Computing (CSC) for
providing computer facilities.

%
% \vspace*{0.3cm}
% \noindent
% \textbf{Competing Interests.}
% The authors do not declare any conflicts of interest, and there is no
% financial interest to report.

% \vspace*{0.3cm}
% \noindent
% \textbf{Authors Contributions.}
%
% %%%%%%%
% \noindent
% \textcolor{red}{\textbf{IZ}: TBA}.\textcolor{red}{\textbf{MN}: TBA}.
% \textcolor{red}{\textbf{GG}: TBA}.
% \textcolor{red}{\textbf{J\v{S}}: TBA}.
% \textcolor{red}{\textbf{FV}: TBA}.
% \textbf{NJM}:
% Conceptualization; Methodology;
% Analysis; Writing -- review \& editing.
% \textbf{MMM}: Simulations; Data processing and analysis.
% \textbf{GRL}: Simulations; Data processing and analysis.
% \textbf{JRS}: Simulations and analysis.
% \textbf{AVK}: Conceptualization; Methodology; Simulations and
% analysis; Writing -- review \& editing.
% \textbf{AVS}: Project administration; Conceptualization;
% Methodology, Simulations and analysis; Writing -- review \& editing.
%
% All authors reviewed the final manuscript.

% % %%%%%%%%%%%%%%%%%%%%%%
% \appendix
%
% %%%%%%%%%%%%%%%%%%%%%%%%%%%%%%%%%%%%%%%%%%%%%%%%%%%%%%%%%%%%%%%%%
% \section{\textcolor{red}{TBA} \label{Appendix01}}

%%%%%%%%%%%%%%%%%%%%%%%%%%%%%%%%%%%%%%
\section*{References}
%%%%%%%%%%%%%%%%%%%%%%%%%%%%%%%%%%%%%%

\end{document}